\documentclass[fleqn,twoside]{article}
\usepackage{espcrc2,epsfig}

\usepackage{amsmath,amstext,amsfonts,amsbsy,amssymb,amscd,bbm}

\usepackage{graphicx}
\usepackage[figuresright]{rotating}

\newcommand{\Dslash}{\relax{\kern+.25em / \kern-.70em D}}

\newcommand{\MSbar}{\overline{\rm MS}}

\newcommand{\be}{\begin{equation}}
\newcommand{\ee}{\end{equation}}
\newcommand{\bd}{\begin{displaymath}}
\newcommand{\ed}{\end{displaymath}}
\newcommand{\bea}{\begin{eqnarray}}
\newcommand{\eea}{\end{eqnarray}}
\newcommand{\ba}{\begin{array}}
\newcommand{\ea}{\end{array}}

\newcommand{\Real}{\relax{\mathsf{\Gamma\kern-.35em R}}}
\newcommand{\Int}{\relax{\mathsf{Z\kern-.40em Z}}}

\newcommand{\cO}{{\cal O}}

\title{\vspace{-4.0cm}
       \rightline{\normalsize ROM2F/2002/19}
       \vspace{-0.1cm}
       \rightline{\normalsize MS-TP-02-4}
       \vspace{-0.1cm}
       \rightline{\normalsize CERN-TH/2002-222}
       \vspace{-0.1cm}
       \rightline{\normalsize September 2002}
       \vspace{2.0cm}
Non-perturbative scale evolution of four-fermion operators\thanks{Work supported
in part by the European Community Human Potential
Programme under contract HPRN-CT-2000-00145, Hadrons/Lattice QCD.}}

\author{ALPHA Collaboration\\
        M.~Guagnelli\address[RM2]{I.N.F.N., Sezione di Roma 2,
        c/o Dipartimento di Fisica, Universit\`a di Roma ``Tor Vergata'',\\ 
        Via della Ricerca Scientifica 1, I-00133 Rome, Italy},
        J.~Heitger\address{Westf\"alische Wilhelms-Universit\"at M\"unster,
        Institut f\"ur Theoretische Physik,\\
        Wilhelm-Klemm-Strasse 9, D-48149 M\"unster, Germany},
        C.~Pena\addressmark[RM2],
        S.~Sint\address{CERN, Theory Division, CH-1211 Geneva 23,
        Switzerland}
        and
        A.~Vladikas\addressmark[RM2]\thanks{Contribution at `Lattice 02', Boston, June 24-29, 2002.}}

\begin{document}

\begin{abstract}
We apply the Schr\"odinger Functional (SF) formalism to determine the renormalisation group running
of four-fermion operators which appear in the effective weak Hamiltonian of the Standard Model.
Our calculations are done using Wilson fermions and the parity-odd components of the operators.
Preliminary results are presented for the operator
$O_{VA} = (\bar s \gamma_\mu d) \,\, (\bar s \gamma_\mu \gamma_5 d)$.
\vspace{1pc}
\end{abstract}

\maketitle

\section{Introduction}

Weak matrix elements (WMEs) of parity-odd four-fermion operators are crucial in many Standard
Model processes (e.g. $\Delta I=1/2$ enhancement, $\Delta S=1$
non-leptonic Kaon decays). The non-perturbative
aspects of these decays are expressed as WMEs $\langle \pi \pi \vert
O_k \vert K \rangle$ of parity-odd operators $O_k$ (where, in standard
notation, $k= VA \pm AV, SP \pm PS, T \tilde T$ denotes the various Dirac structures).
On the other hand, the non-perturbative aspects of $K^0 - \bar K^0$
oscillations (i.e. $\Delta S=2$ transitions) are studied through WMEs
of parity-even operators $\langle \bar K^0 \vert O_{VV+AA} \vert K^0
\rangle$. Even in this case, however, it is possible to map the problem
into the computation of parity-odd matrix elements, either through the
use of Ward Identities~\cite{becetal} or by implementing
the twisted mass QCD formalism~\cite{tmQCD}.

Unlike their parity-even counterparts, all parity-odd operators of
interest display
renormalisation patterns which are unaffected by the breaking of
chiral symmetry on the lattice (i.e. the Wilson term)~\cite{Bernard}.
Our program consists in computing non-perturbatively the anomalous
dimension (AD) matrix of the complete basis of parity-odd dimension-six operators with
Wilson fermions. Our renormalisation scheme of choice is the SF, which
is mass independent. This has two implications. First, our results also apply to
those operators which characterise $\Delta I =1/2$
transitions. Although the complete removal of divergences in this case
requires complicated power subtractions, these, being proportional to quark masses,
do not affect the operator AD. Second, our results also
apply to operators of heavy flavour transitions (e.g $\Delta C,\Delta
B = 2,~\Delta B = 1$).

The renormalisation of any four-fermion operator (up to power
subtractions) can be mapped to that of an operator with the same
colour-Dirac structure but four distinct flavours. For example, once the operators
\begin{align}
        O_{VA+AV}^\pm  = \frac{1}{2} \big[
                          \big( &\bar \psi_1 \gamma_\mu \psi_2\big)
                           \big(\bar \psi_3 \gamma_\mu \gamma_5 \psi_4\big) + \nonumber\\
                        + \big(\bar \psi_1 \gamma_\mu \gamma_5 \psi_2 &\big)
                           \big(\bar \psi_3 \gamma_\mu \psi_4\big)
                        \pm (2 \leftrightarrow 4) \big ]
\end{align}
have been properly renormalised, we can map the four distinct
flavours to the physical flavours concerning the process of
interest. Here we will only present results for 
the operator $Q \equiv O_{VA+AV}^+$.

\section{The step scaling function (SSF)}

Our work consists in a straightforward generalisation of the
computation of the SSF of the quark mass, performed by the Alpha
collaboration in~\cite{alpha_mq}. The operator anomalous
dimension $\gamma$ is defined through
\bea
\mu \,\,\, \frac{\partial \bar Q}{\partial \mu} =
\gamma(\bar g) \,\,\, \bar Q
\label{eq:rg}
\eea
($\bar Q$ and $\bar g$ are renormalised quantities).
Its perturbative expansion is given by
\bea
\gamma(g) \stackrel{g \to 0}{\sim} -g^2\left(\gamma_0 + \gamma_1 g^2 +
\gamma_2 g^4 + \ldots \right) \ ,
\eea
with $\gamma_0 = 1/(4 \pi^2)$ a universal coefficient. The (scheme-indepedent) renormalisation group invariant (RGI)
operator is obtained upon integrating eq.~(\ref{eq:rg})
\begin{align}
Q^{RGI} \,\, =& \,\, \bar Q(\mu) \,\,\, [2 b_0 \bar g^2]^{(-\gamma_0/2b_0)} \nonumber \\
\times \exp&\Big\{ - \int_0^{\bar g} dg \left[ 
\frac{\gamma(g)}{\beta(g)} - \frac{\gamma_0}{b_0 g} \right] \Big\}
\label{eq:rgi}
\end{align}
($b_0$ is the LO coefficient of the Callan-Symanzik $\beta$-function).\
Once the scale dependence of the renormalised coupling $\bar g$ in the SF scheme is known~\cite{alpha_g},
and proceeding along the line of ref.~\cite{alpha_mq},
three elements are required for the complete NP determination of a WME of the form
$\langle f \vert Q^{RGI} \vert i \rangle$:
(i) the lattice bare matrix elements at several bare couplings (work in progress);
(ii) the value of the renormalisation constant $Z_Q$ at the same bare
couplings and a fixed low-energy reference scale of the order of $\Lambda_{QCD}$;
(iii) the operator SSF from this non-perturbative low-energy scale to high ones,
where $\bar g(\mu)$ is safely in the perturbative region and the integral in eq.~(\ref{eq:rgi})
can be performed analytically to determine the ratio $Q^{RGI}/\bar Q(\mu)$.

\begin{figure}[htb]
\vspace{2.0cm}
\includegraphics{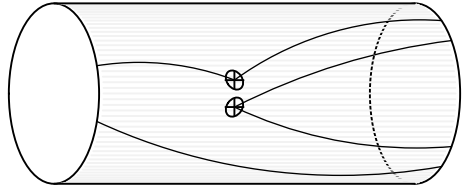}
\caption{
The SF correlation function $h_Q$.
Note that the ``spectator'' quark line connecting the two time
boundaries corresponds to a fifth flavour.
}
\label{fig:hq}
\end{figure}

We regularise the theory on a lattice of physical size $L^4$ with
standard SF boundary conditions~\cite{alpha_mq}.
Renormalisation is performed in the chiral limit, at a scale equal
to the IR cutoff $L^{-1}$.
Since the operator renormalises multiplicatively,
\bea
\bar Q (L^{-1}) = \lim_{a\to 0} \,\,\, Z_Q(g_0,L/a) \,\,\, Q (a) \ ,
\eea
the UV divergence is removed by a single renormalisation condition,
namely by setting the expression
\bea
f_1^{-3/2}(g_0,L/a) \,\,\, Z_Q(g_0,L/a) \,\,\, h_Q(x_0)
\eea
at $x_0 = L/2$ to its tree level value.
The correlator $h_Q$ is as shown in Fig.~\ref{fig:hq}; $f_1$
serves to cancel the quark boundary field renormalisation (see
ref.~\cite{alpha_mq}).  For more details and an explicit notation, see
ref.~\cite{alpha_Q}.
Different choices of
boundary operators correspond to different renormalisation schemes.
Parity conservation and the properties of SF boundary quark fields
limit us to five non-trivial independent choices, 
\begin{align}
\gamma_5 \leftrightarrow \gamma_5 \,\, \gamma_5 \nonumber \quad & \quad
\gamma_5 \leftrightarrow \gamma_k \,\, \gamma_k \nonumber \\
\gamma_k \leftrightarrow \gamma_k \,\, \gamma_5 \nonumber \quad & \quad
\gamma_k \leftrightarrow \gamma_5 \,\, \gamma_k \nonumber \\
\gamma_i \leftrightarrow \gamma_j &\,\, \gamma_k \,\, \epsilon_{ijk} \ ,
\label{eq:bounds}
\end{align}
plus their time-reversed counterparts (in eq.~(\ref{eq:bounds}) the
$\gamma$-matrix on the lhs. of the symbol $\leftrightarrow$ corresponds to
the Dirac structure at time $x_0 = 0$, whereas the two
$\gamma$-matrices on its rhs. correspond to the Dirac structures at
time $x_0 = L$).

The SSF is defined as
\bea
\sigma_Q(u) \equiv \lim_{a \to 0} \Sigma_Q(u,a/L) \equiv \lim_{a \to 0} \frac{Z_Q(g_0,2L/a)}{Z_Q(g_0,L/a)} \relax{\kern-2mm}
\eea
($u = \bar g^2(L)$) and, taking into account the running coupling from~\cite{alpha_g}, can be used
to run renormalised matrix elements along a broad range of scales through the recursion
\bea
\bar Q((2L)^{-1}) = \sigma_Q(u) \bar Q(L^{-1}) \ .
\eea
\begin{figure}[h!]
\vspace{4.8cm}
\includegraphics{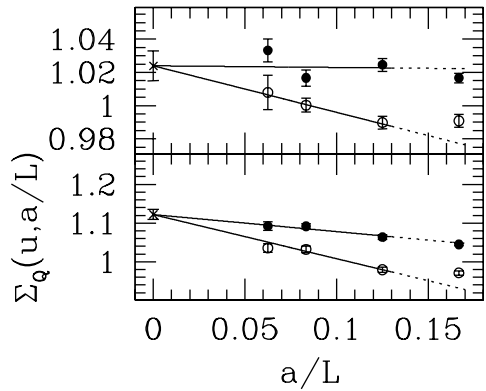}
\caption{
Continuum limit extrapolation of the SSF at renormalised couplings
$u=1.2430$ (upper plot) and $u=2.770$ (lower plot). Solid (open) points are data corresponding to simulations
with an improved Clover (Wilson) action. The leftmost point in each plot is the result of a combined
linear extrapolation to the continuum limit of both data sets, with slopes indicated
by the two lines. In both cases, the data at largest lattice spacing have not been
included in the fit.
}
\label{fig:extrap}
\end{figure}
Our quenched simulations have been perfomed at the parameters
(volumes, couplings and hopping parameters) of ref.~\cite{alpha_mq}.
The most difficult numerical task consists in the extrapolation of our
data to the continuum limit: although our action is the
non-perturbatively improved Clover one (i.e. cutoff effects are $\cO(a^2)$ from the
bulk, and $\cO(g_0^4 a)$ from the time boundaries), our operator is
unimproved, which gives rise to $\cO(a)$ effects. Thus, for some 
renormalised couplings the continuum limit extrapolation appears to be stable,
whereas for others it is rather unreliable.
In order to improve on this we are currently repeating
our computation, in the spirit of ref.~\cite{gjp}
with an unimproved Wilson action. Since the SSF
corresponding to the same renormalised coupling has the same continuum
limit, combined fits of the Clover and Wilson SSF should stabilise the
results. To show that this is actually the case, we present in 
Fig.~\ref{fig:extrap}
two of our first combined extrapolations at two different values
of the renormalised coupling.
Until this analysis is completed, we shall refrain from quoting
an estimate for the quantity $Q^{RGI}/\bar Q(\mu)$.

\begin{figure}
\vspace{5.4cm}
\includegraphics{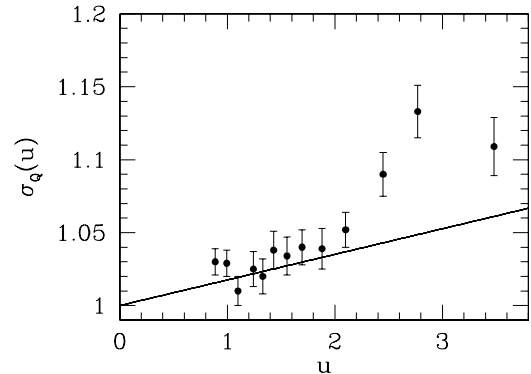}
\vspace{-0.8cm}
\caption{
Step scaling function $\sigma_Q$ in the continuum limit. The solid line is the
LO perturbative prediction for $\sigma_Q$.
}
\label{fig:ssf}
\end{figure}

In Fig.~\ref{fig:ssf} we show our present results with an improved action for the SSF
associated with the first renormalisation condition in~eq.~(\ref{eq:bounds}).
The step scaling functions associated with the other choices of renormalisation
scheme are qualitatively similar.
Our results need to be supplemented by the NLO calculation of the SSF,
which will enable us to estimate the importance of NP contributions to
this quantity in the SF scheme. It will also provide the necessary
information for the matching between SF and $\MSbar$ schemes.
This calculation is well under way.

\section*{Acknowledgements}
We acknowledge useful discussions with M.~Della~Morte, R.~Frezzotti, K.~Jansen,
and R.~Sommer.

\end{document}